\documentclass[aps,prb,reprint,amsmath,amssymb,superscriptaddress,notitlepage]{revtex4-2}
\usepackage[utf8]{inputenc}
\usepackage{bm}
\usepackage{enumerate}
\usepackage{graphicx}
\usepackage{cancel}
\usepackage{xcolor}
\usepackage{verbatim}
\usepackage{hyperref}

\hypersetup{colorlinks=true,
linkcolor=blue,
citecolor=blue,
urlcolor=blue,
filecolor=blue
}

\begin{abstract}

Electronic devices that work in the quantum regime often employ hybrid nanostructures to bring about a nonlinear behaviour. The nonlinearity that these can provide has proven to be useful, in particular, for applications in quantum computation. %
Here we present a hybrid device that acts as a capacitor with a nonlinear charge-voltage relation. The device consists of a nanowire placed between the plates of a coplanar capacitor, with a co-parallel alignment. At low temperatures, due to the finite density of states on the nanowire, the charge distribution in the capacitor is uneven and energy-dependent, resulting in a charge-dependent effective capacitance. %
We study this system analytically and numerically, and show that the nonlinearity of the capacitance is significant enough to be utilized in circuit quantum electrodynamics. The resulting nonlinearity can be switched on, modulated, and switched off by an external potential, thus making this capacitive device highly versatile for uses in quantum computation.
\end{abstract}

\begin{document}

\title{Superconductor-semiconductor hybrid capacitance with a nonlinear charge-voltage profile}

\author{Joachim Lauwens}
\affiliation{Imec, Kapeldreef 75, 3001 Heverlee, Belgium.}
\affiliation{Department of Physics and Astronomy, KU Leuven, Celestijnenlaan 200D, 3001 Leuven, Belgium.}

\author{Lars Kerkhofs}
\affiliation{Imec, Kapeldreef 75, 3001 Heverlee, Belgium.}

\author{Arnau Sala}
\altaffiliation{Current address: JARA-FIT Institute for Quantum Information, Forschungszentrum J\"ulich GmbH and RWTH Aachen University, 52074 Aachen, Germany}
\affiliation{Imec, Kapeldreef 75, 3001 Heverlee, Belgium.}
\affiliation{Department of Materials Engineering, KU Leuven, Kasteelpark Arenberg 44, 3001 Leuven, Belgium.}

\author{Bart Sor\'ee}
\affiliation{Imec, Kapeldreef 75, 3001 Heverlee, Belgium.}
\affiliation{Department of Electrical Engineering, KU Leuven, Kasteelpark Arenberg 10, 3001 Leuven, Belgium.}
\affiliation{Department of Physics, University of Antwerp, Groenenborgerlaan 171, 2020 Antwerp, Belgium.}

\date{\today}

\maketitle

\section{Introduction}

Nonlinear devices play a key role in circuit quantum electrodynamics (cQED) by e.g., introducing anharmonicities in an energy spectrum~\cite{Devoret1995,krantz2019quantum,blais2021circuit}, enabling different forms of interactions~\cite{moskalenko2022high, aumentado2020superconducting} and allowing for extra electronic~\cite{casparis2018superconducting,Larsen2015} or magnetic~\cite{kounalakis2018tuneable} tunability of various Hamiltonian parameters~\cite{Materise2022, Ilani2006}.

The properties of such devices often arise from a nonlinear current-phase relation, as in a Josephson tunnel junction~\cite{Nakamura1999,koch2007charge}, or from a nonlinear charge-voltage relation, as in semiconductor-superconductor hybrid capacitive devices~\cite{Ilani2006,khorasani2017nonlinear,Balasubramanian2021}. 

Capacitive devices with a nonlinear charge-voltage regime have been studied in detail and are broadly used for different applications in conventional electronics, such as zero-voltage switches, snubber circuits or high-voltage pulse generation~\cite{Drofenik2010,Hakim2000,Rim1999}. Two of the most prominent examples of such nonlinear devices are the varicap, which consists of a reverse-biased p-n junction used as a voltage-controlled capacitor~\cite{Hum2005,Mortenson1974,Penfield1962}, and the metal-oxide-semiconductor capacitor (MOScap), which has found applications in low electric noise amplification, high-speed optical phase modulation or for the detection of gases~\cite{Figueiredo2004,Liu2004,Medlin2003,Soo2010}.

Examples of nonlinear capacitive devices operating at low temperatures and low voltages---the regime of operation of most qubit implementations---are more sparse, as these often involve complex optomechanical~\cite{aspelmeyer2014cavity,toth2017dissipative} or multilayered~\cite{Ilani2006,Balasubramanian2021,khorasani2017nonlinear} structures that limit their potential for scalability.

\begin{figure}
    \centering
    \includegraphics[width=\linewidth]{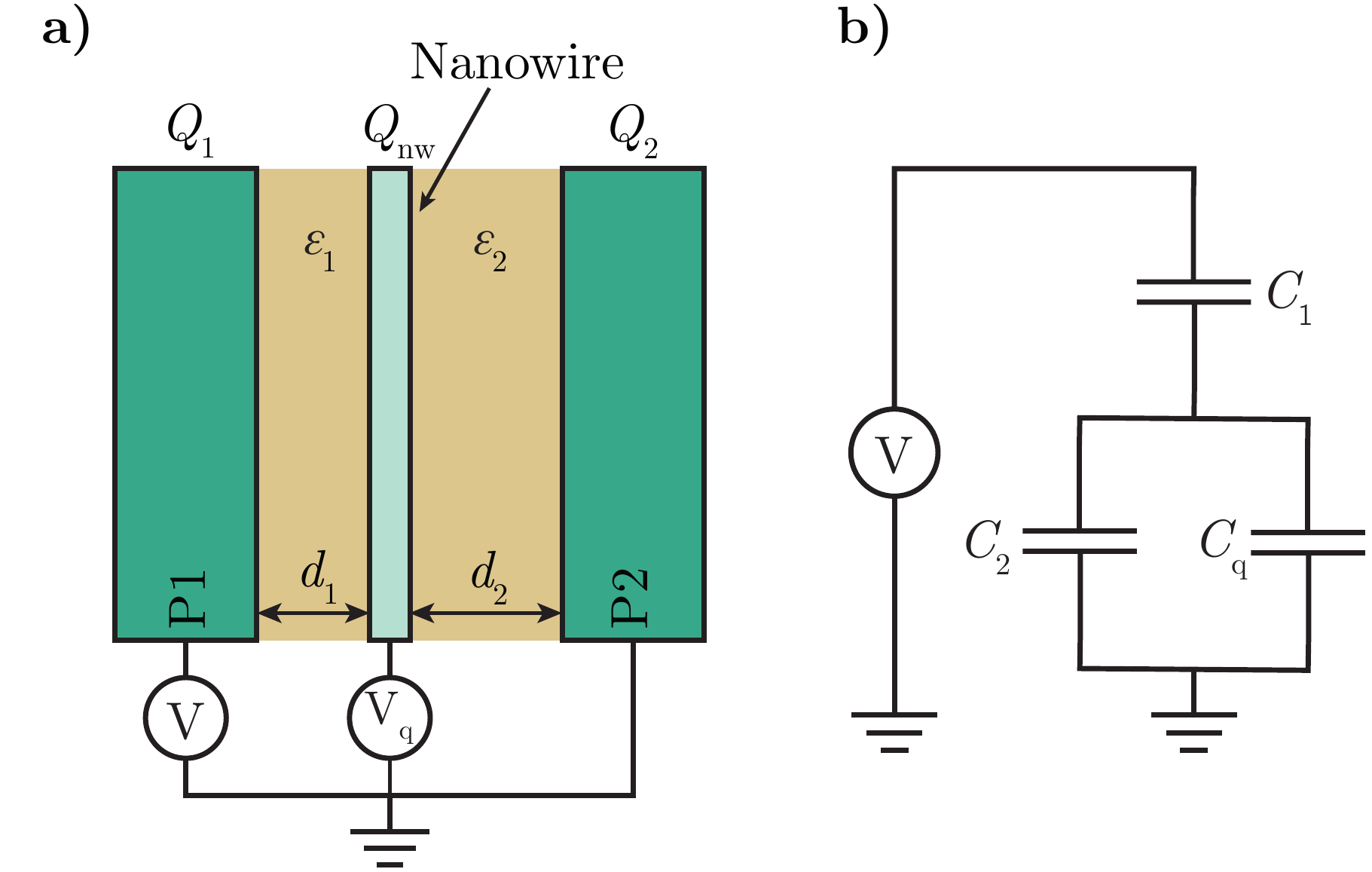}
    \caption{\textbf{a)} Schematic of the device. \textbf{b)} Equivalent electrical circuit of the device. $C_1$ and $C_2$ are the geometric capacitances between the plates and the nanowire, $C_q$ is the quantum capacitance due to the low density of states of the nanowire.}
    \label{fig:device}
\end{figure}

Here we investigate a simple, two-dimensional capacitive device that displays a nonlinear charge-voltage relation in the regime of operation of cQED. This planar structure, schematically depicted in Fig.~\ref{fig:device}, consists of a one-dimensional electron gas (1DEG) in a nanowire positioned between two superconducting plates. Due to the small density of states (DOS) of the 1DEG, an incomplete screening of the charge accumulated on the plates by the electron gas~\cite{luryi1988quantum,John2004,Akinwande2007,Fang2007,Balasubramanian2021} leads to a nonlinear, voltage-dependent capacitance that can be used in conjunction with other cQED elements for applications in quantum computation and quantum parametric amplification~\cite{Devoret1995,sala2016proposal,Roy2016}.
The gatemon, a similar device composed of the same elements but in a different geometry, have been investigated, fabricated and successfully implemented in a cQED setup~\cite{casparis2018superconducting,Larsen2015}.  While similar fabrication techniques may be used, the main difference between our device and the gatemon is that in our device the distance between the plates and nanowire is large enough to prevent tunneling of charge carriers, thus acting as a purely capacitive device. We present an analytical and numerical study of this nonlinear capacitive device, taking into account various band structures and thermal effects.

This paper is organised as follows: In section~\ref{analytical model} we introduce a simple, analytical model of the device at zero temperature and for a parabolic 1DEG band structure; in section~\ref{numerics} we generalize the model to non-zero temperatures and more complex band structures, and calculate the capacitance numerically; next, in section~\ref{applications} we discuss the possibility to implement the nonlinear capacitor in a cQED circuit for applications in quantum computation.
 
\section{Analytical model at T=0 K}
\label{analytical model}

We study a device composed of two superconducting, coplanar, parallel plates, and a nanowire. The nanowire is placed between the plates, has the same length and is co-parallel to the plates, as depicted in Fig.~\ref{fig:device}. For the sake of simplicity, we consider plates of infinite length and finite width.
The plate on the left, labeled P1, is connected to a potential $V$ and accumulates a charge per unit length $Q_1$, while the plate on the right, P2, is grounded and has a charge per unit length $Q_2$. The plates and the nanowire are separated by dielectrics with dielectric constants $\varepsilon_1$ and $\varepsilon_2$ for the left and right dielectrics, respectively.
We refer to the constant geometric capacitances per unit length resulting from the dielectrics $\varepsilon_1$ and $\varepsilon_2$ as $C_1$ and $C_2$.
The role of the nanowire is to host a 1DEG along the length of the device. We thus consider a donor doping such that the Fermi level of the nanowire lays at the bottom of the band minimum~\cite{Funk2013, Vezzosi2022}. The Fermi level can then be raised and lowered by applying a potential $V_\text{q}$ to the 1DEG. This control over the Fermi level directly provides an electrical tunability of the nonlinear behaviour of the device, by ofsetting the charge density in the nanowire.

The carrier density of the nanowire, defined as $n\equiv -Q_\text{nw}/e$, where $e$ is the elementary charge, is given by the Fermi integral

\begin{align}\label{eq:charge_intergal}
    n(\mu, T) ={}&{} \int_{-\infty}^{+\infty} f(E, T, \mu)\, g(E)\, dE,
\end{align}

where $f(E, T, \mu)$ is the Fermi-Dirac distribution at temperature $T$, and $g(E)= g_0/\sqrt{E}$ is the density of states in the conduction band of the nanowire. For a 1DEG with a single, perfectly parabolic band and at zero temperature, the integral above has an analytical solution, namely $n(\mu, T=0) = 2 g_0 \sqrt{\mu - U_\text{q}}$. Here $g_0 = g_v \sqrt{2 m_e}/\pi\hbar$ is a material-dependent constant, where $g_v$ is the valley degeneracy of the semiconductor and $m_e$ is the effective electron mass in the conduction band. In this expression we have also included a shift in the chemical potential $U_\text{q}$ resulting from the externally applied potential $V_\text{q}$. For the sake of simplicity, we define the potential $V_\text{q}$ relative to this energy shift, as $U_\text{q}\equiv -e V_\text{q}$. Similarly, the Fermi degeneracy energy can be evaluated using the integral

\begin{align}\label{eq:energy_integral}
    E_\text{nw}(\mu, T) ={}&{} \int_{-\infty}^{+\infty} E\, f(E, T, \mu)\, g(E)\, dE.
\end{align}

By combining Eqns.~(\ref{eq:charge_intergal}) and~(\ref{eq:energy_integral}), we can express the Fermi degeneracy energy as a function of the charge accumulated in the nanowire:

\begin{align}\label{eq:energy_nw}
    E_\text{nw} = -\frac{Q_\text{nw}^3}{12 g_0^2 e^3} + Q_\text{nw} V_\text{q}.
\end{align}

The total energy of the system is then given by the individual contribution of the electric field energies from the two dielectrics and the Fermi degeneracy energy of Eq.~(\ref{eq:energy_nw}). That is,
\begin{align}
    E_\text{tot} = \frac{Q_1^2}{2 C_1} + \frac{Q_2^2}{2 C_2} - \frac{Q_\text{nw}^3}{12 g_0^2 e^3} + Q_\text{nw} V_\text{q}.
    \label{eq:totE}
\end{align}

By imposing the charge neutrality condition $Q_1 + Q_2 + Q_\text{nw} = 0$ and minimizing the total energy, we obtain the expressions for the charge accumulated in P2 and in the nanowire as a function of $Q_1$, for $Q_1 > 0$:

\begin{align}
    Q_2 ={}&{} - Q_1 - \frac{2 e^3 g_0^2}{C_2} \left[ 1 - \sqrt{1 + \frac{C_2 (Q_1 + C_2 V_\text{q})}{e^3 g_0^2}} \right], \label{eq:q2} \\
    Q_\text{nw} ={}&{} \frac{2 e^3 g_0^2}{C_2} \left[ 1 - \sqrt{1 + \frac{C_2 (Q_1 + C_2 V_\text{q})}{e^3 g_0^2}} \right]. \label{eq:qnw}
\end{align}

For $Q_1 < 0$, the Fermi level of the 1DEG is pushed towards the semiconductor gap, where the DOS is exactly zero at $T=0$\,K. In this scenario, we have $Q_2 = -Q_1$ and $Q_\text{nw} = 0$. 

\begin{figure}[t]
    \centering
    \includegraphics[width=\linewidth]{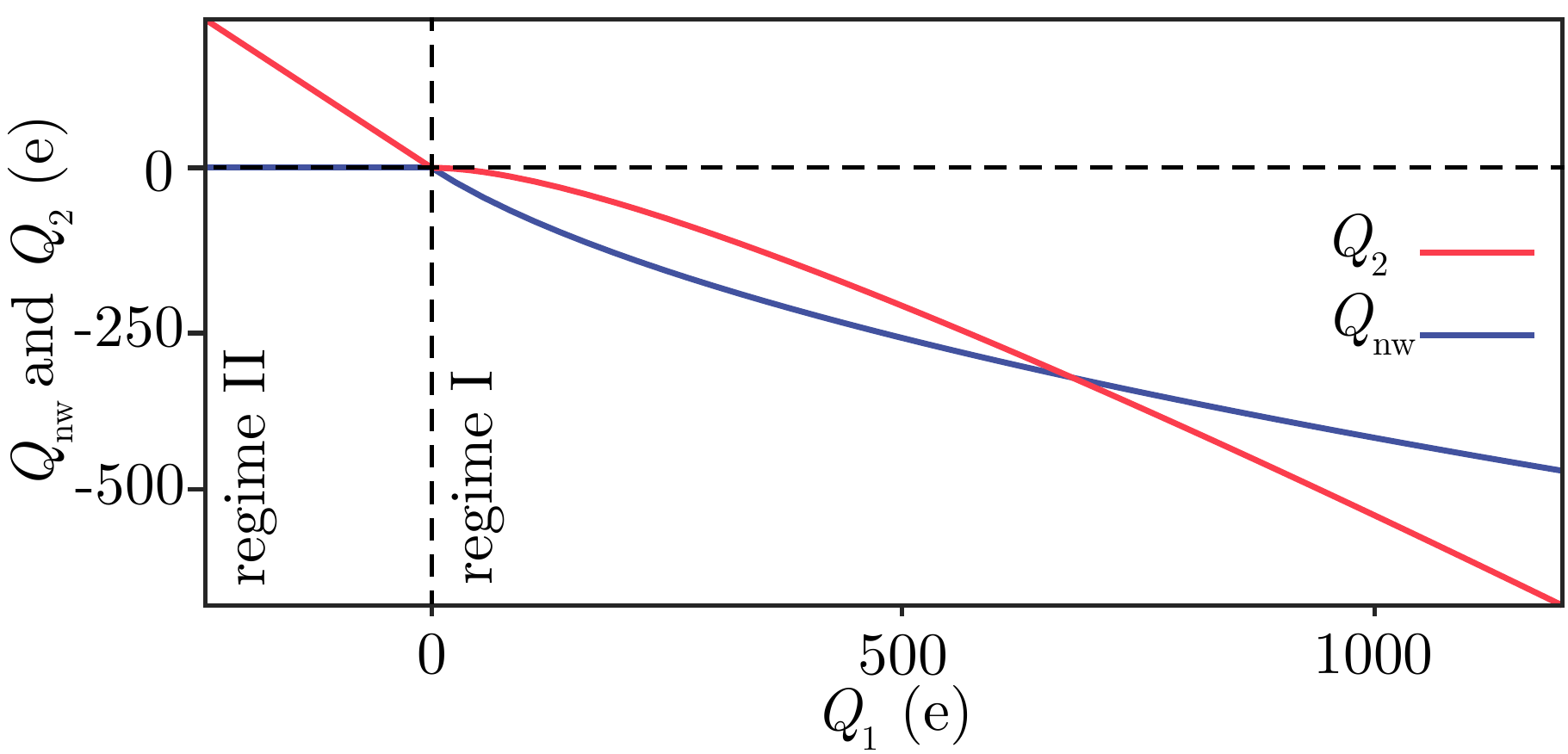}
    \caption{Accumulated charge $Q_2$ and $Q_\text{nw}$ plotted as a function of $Q_1$ for $V_\text{q}=0$, according to Eqns.~(\ref{eq:q2})-(\ref{eq:qnw}). At $Q_1 = 0$ there is a clear transition from the linear to the nonlinear regime, as more charges occupy the nanowire. The parameters used in the plot are $C_1 = 10\,$fF/$\mu$m, $C_2 = 0.15\,$fF/$\mu$m,  $m_e = 0.067 m_0$, with $m_0$ the free electron mass, and a device length of $1\, \mu$m.}
    \label{fig:charges}
\end{figure}

We show the charge filling of the nanowire and the plate P2 as a function of the charge $Q_1$ in Fig.~\ref{fig:charges}, where we have chosen $V_\text{q}=0$. For negative $Q_1$, the nanowire cannot accumulate any charge, and the device is in the linear regime, where the charge in opposite plates has the same absolute value with a different sign. As the plate P1 is filled with a positive charge, the nanowire starts filling with a negative one, but due to the only partial screening of the electric field from P1, soon the plate P2 starts filling as well with a negative charge. When the charge $Q_1$ reaches the critical value $Q_1 = 8 e^3 g_0^2/C_2$, the filling of the nanowire and P2 are the same. After this point, the nanowire continues increasing the negative charge following a square root, while P2 tends asymptotically to a linear regime.

We evaluate the potential across the plates P1 and P2 by taking the difference between the potential created by a charge $Q_1$ on P1 and the potential created by a charge $Q_2$ on P2, resulting in

\begin{align} \label{eq:v_tot}
    V = {}&{} \left( \frac{1}{C_1} + \frac{1}{C_2} \right) Q_1 \notag \\
    {}&{} + \frac{2 e^3 g_0^2}{C_2^2} \left[ 1 - \sqrt{1 + \frac{C_2 (Q_1 + C_2 V_q)}{g_0^2 e^3}} \right].
\end{align}

The total capacitance of the system is then defined as $C = dQ_1/dV$. In the regime I of Fig.~\ref{fig:charges}, where the plate P1 is filled by a positive charge, this is

\begin{align} \label{eq:c_tot}
    C_\text{I}(Q_1) = \frac{1}{\frac{1}{C_1} + \frac{1}{C_2 + C_\text{q}(Q_1)}},
\end{align}

with $C_\text{q}$, given by

\begin{align} \label{eq:c_q}
    C_\text{q}(Q_1) = \frac{C_2}{\sqrt{1 + \frac{C_2 (Q_1 + C_2 V_\text{q})}{g_0^2 e^3}} - 1}.
\end{align}

The form of the total capacitance in Eq.~(\ref{eq:c_tot}) indicates that the limited density of states of the nanowire increases the total capacitance of the system. This increase of the total capacitance can be seen as the result of the addition of an extra capacitor, with capacitance $C_\text{q}$, as depicted in the equivalent circuit of Fig.~\ref{fig:device}b.
This quantity is commonly referred to as the quantum capacitance, and it arises when the charge accumulated in one of the components of a capacitor is limited by a small DOS~\cite{luryi1988quantum}.
Furthermore, as a consequence of the energy dependence of the DOS of the 1DEG in the nanowire, this quantum capacitance depends on the charge accumulated in the plates. Or, equivalently, on the potential across the plates, if we invert Eq.~(\ref{eq:v_tot}) and express Eq.~(\ref{eq:c_q}) as a function of $V$.

\begin{figure}[t]
    \centering
    \includegraphics[width=\linewidth]{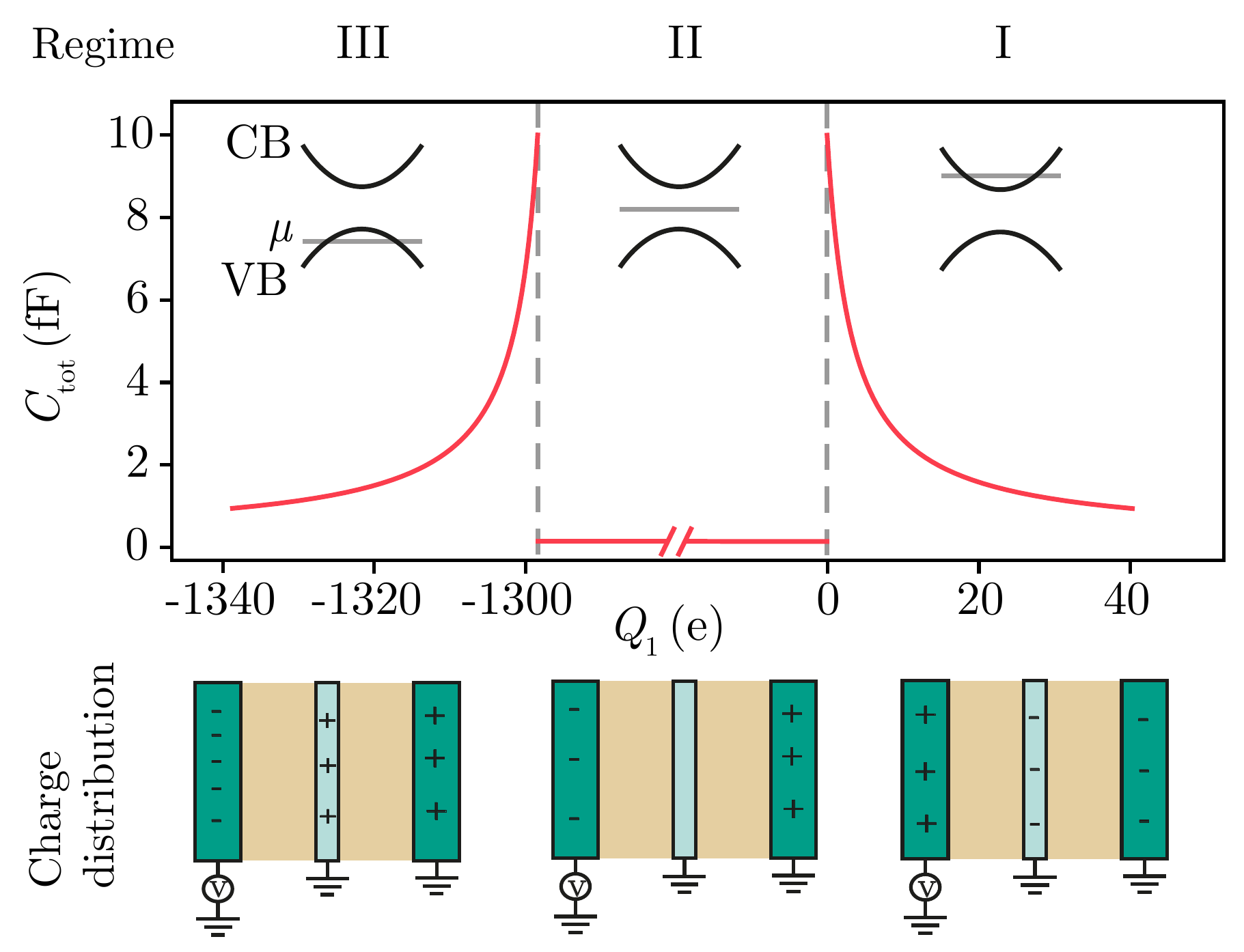}
    \caption{Operating regimes of the nonlinear capacitor. The total capacitance shows three different regimes: (I) when the Fermi level rises above the conduction band, the total capacitance decreases with increasing charge $Q_1$ in P1; (II) when the Fermi level lies inside the band gap of the nanowire, the total capacitance is constant, we call this the linear regime; (III) pushing down the Fermi level below the valence band maximum takes the device to the third regime, where the 1DEG is filled with holes and the total capacitance decreases with decreasing charge $Q_1$. The panel below shows the charge distribution in P1, P2 and the 1DEG. This figure has been obtained by evaluating the total capacitance analytically as indicated in the main text, for the three different regimes separately. Here we have used $m_e = m_h = 0.067 m_0$. Additionally, we have used $T=0$, $C_1 = 10.0\,$ fF/$\mu$m and $C_2 = 0.15\,$ fF/$\mu$m.}
    \label{fig:regimes}
\end{figure}

We show the behavior of the total capacitance as a function of the charge $Q_1$ in Fig. \ref{fig:regimes}, for zero temperature and $V_\text{q}=0$.
When the Fermi level rises above the bottom of the conduction band, that is, for $Q_1 + C_2 V_q \geq 0$, the nanowire starts filling with excess electrons. This is the regime of validity of Eq.~(\ref{eq:c_tot}). In the vicinity of $Q_1=0$, virtually all the charge in $Q_1$ is screened by the 1DEG (see also Fig.~\ref{fig:charges}), and the capacitor consists only of a plate P1 and a 1DEG. At this point, the total capacitance is thus $C_1$. As the charge $Q_1$ keeps increasing, the total capacitance decreases as a power law towards the limit where the 1DEG cannot screen any more electric charge and the device behaves as a conventional capacitor with plates P1 and P2. In this limit, the total capacitance is then $C_1 C_2/(C_1 + C_2)$.
If, on the other hand, the plate P1 accumulates a negative charge, we then enter in the regime II, where the Fermi level lays within the band gap of the nanowire. In this regime, the 1DEG cannot screen any charge and the total capacitance is again $C_1 C_2/(C_1 + C_2)$.
Further decreasing $Q_1$ brings us to the third regime, where the 1DEG is now filled with excess holes. Assuming a density of states for holes identical to that of the electrons, the total capacitance in this regime mirrors the one in regime I.

\section{Numerical evaluation} 
\label{numerics}

\begin{figure}[t]
    \centering
    \includegraphics[width=\linewidth]{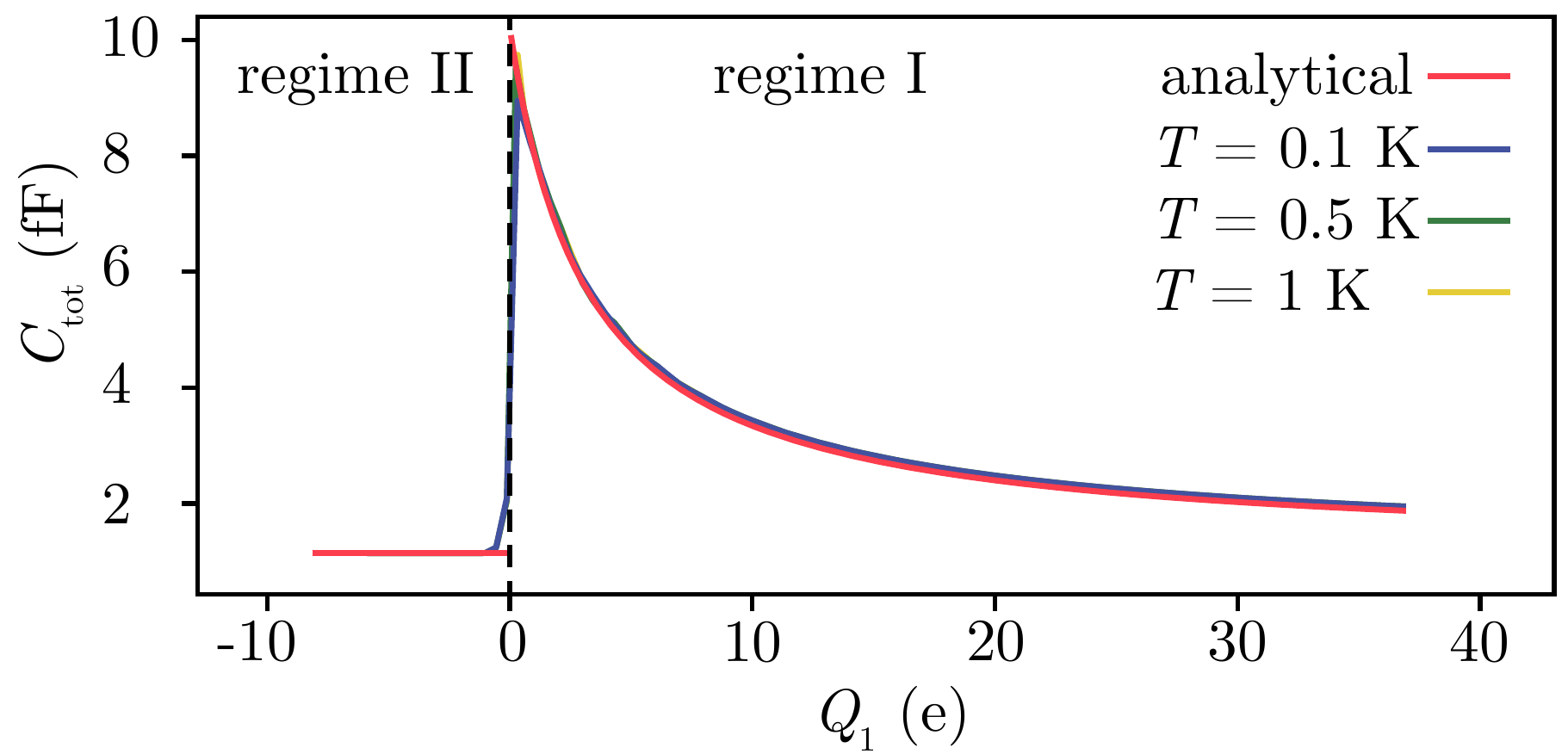}
    \caption{Capacitance as a function of $Q_1$, comparing the analytical approximation at $0$K (red) to numerical results for temperatures of $0.1$K, $0.5$K and $1$K, and the same device parameters as in Fig.~\ref{fig:charges}.}
    \label{fig:temperature}
\end{figure}

For finite temperatures and a more accurate density of states, the solutions of the Fermi integrals of Eqns.~(\ref{eq:charge_intergal}) and ~(\ref{eq:energy_integral}) do not have a closed form. We thus evaluate the total capacitance for $T\neq 0$ and a more general DOS numerically. Since the scope of this paper is the study of a nonlinear device for applications in cQED, we limit the temperature range to a few K~\cite{berns2006coherent}. Fig.~\ref{fig:temperature} compares the calculated total capacitance for temperatures of $T = 0.1$K, $T = 0.5$K and $T = 1$K, to the analytical expression derived in section~\ref{analytical model}. In this range, the effect of the temperature on $C_\text{tot}$ is negligible.
We take into account the effects of the finite size of the 1DEG, electron correlations and the quantum confinement as a deviation from parabolicity of the energy bands of the nanowire and thus use the Kane model for the DOS~\cite{godoy2005effects,Jin2007}. For a single band model, we use

\begin{align}
    g(E) = \frac{g_v\sqrt{2m}}{\hbar\pi}\frac{1+2\alpha E}{\sqrt{E(1+\alpha E)}},
\end{align}

where $\alpha$ is a measure for the degree of the non-parabolicity of the energy band.

\begin{figure}[t]
    \centering
    \includegraphics[width=\linewidth]{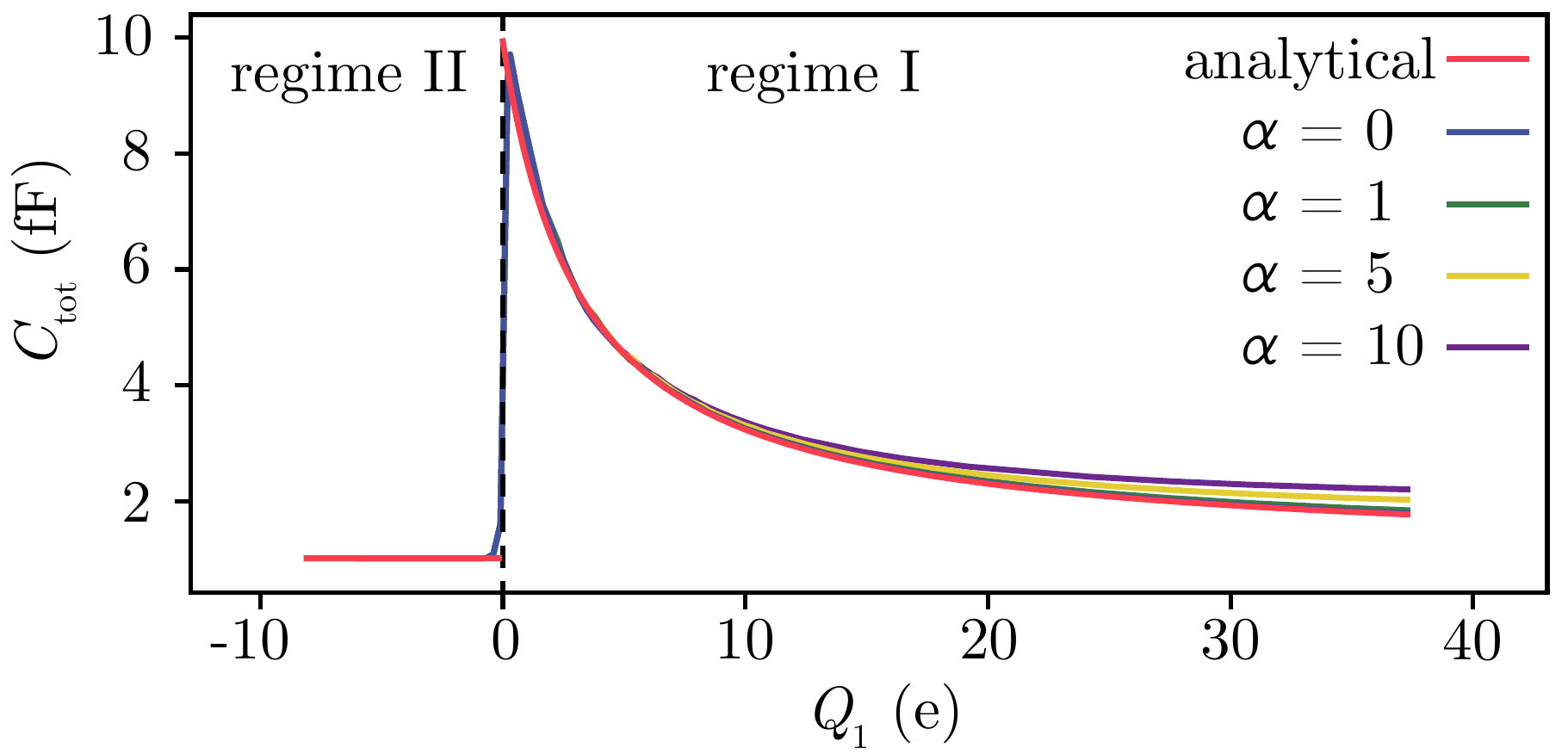}
    \caption{Capacitance as a function of $Q_1$, comparing the analytical approximation at $0$K (red) to numerical results for a temperature of $1$K, with $\alpha$-values ranging from $0$ $\text{eV}^{-1}$ to $10$ $\text{eV}^{-1}$, and the same device parameters as in Fig.~\ref{fig:charges}.}
    \label{fig:nonparabolicity}
\end{figure}

In Fig.~\ref{fig:nonparabolicity} we show the total capacitance for different values of $\alpha$, ranging from 0\,eV$^{-1}$ to 10\,eV$^{-1}$, and for T=1\,K, together with the analytical solution calculated for $\alpha = 0\,$eV$^{-1}$ and $T=0\,$K. In this figure we observe that at low temperatures, deviations from parabolicity of the energy bands only result in an offset of the capacitance at large charge accumulation.
This difference from the limit of $\alpha=0$ results from the behavior of the DOS at large energies, which reaches the constant limit $g(E\to \infty) = 2g_0\sqrt{\alpha}$, whereas for perfectly parabolic bands, $g(E\to \infty) = 0$. In this limit the quantum capacitance converges to the constant value $C_q = 2g_0e^2\sqrt{\alpha}$ instead of zero, causing the shift of the capacitance in regime I.
Increasing the temperature to $T=1\,$K, on the other hand, does not change the capacitance by a considerable amount.

\begin{figure}[t]
    \centering
    \includegraphics[width=\linewidth]{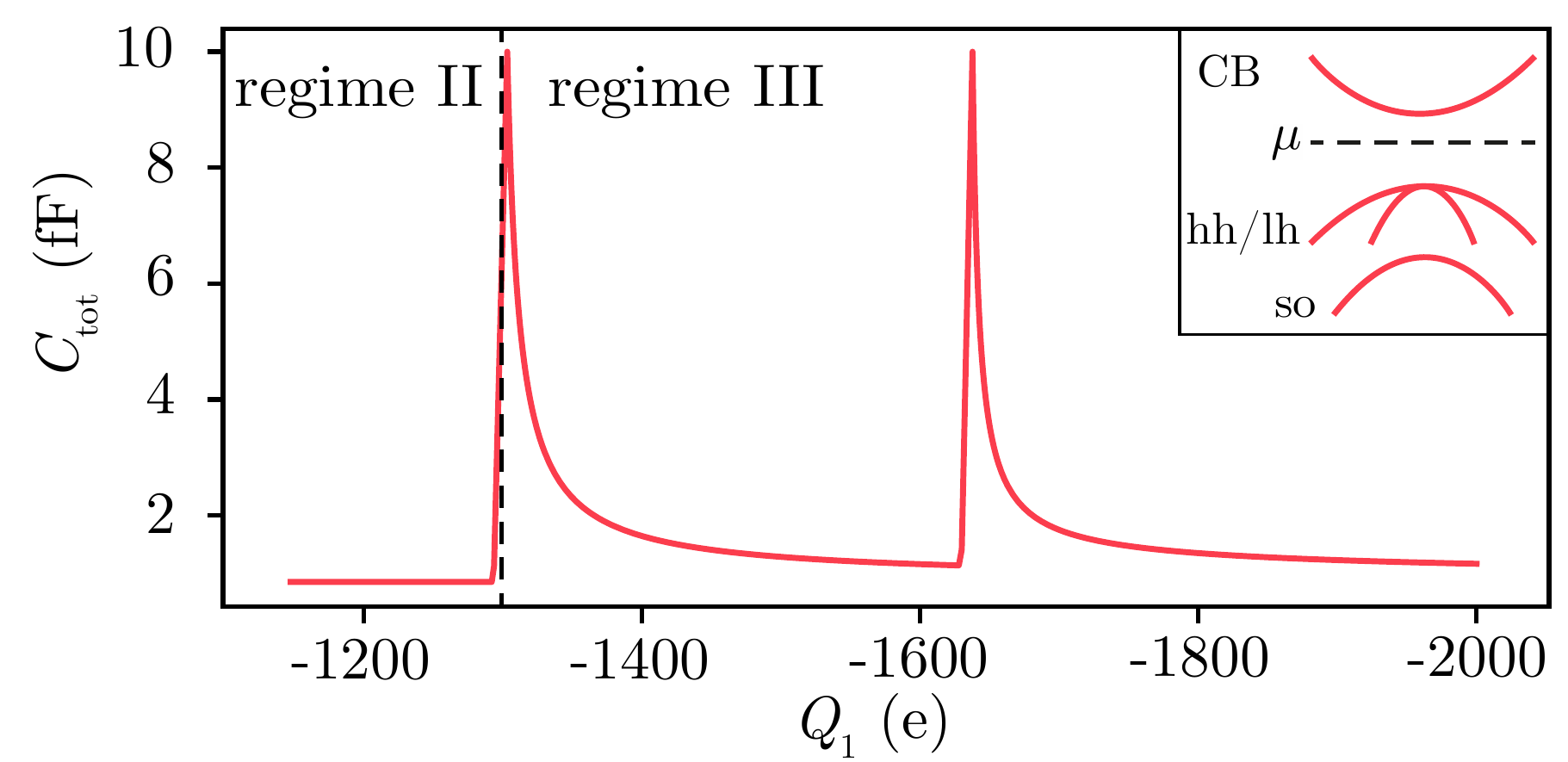}
    \caption{Numerically calculated capacitance as a function of $Q_1$ for a GaAs nanowire, showing the effects of the valence band structure between regimes II and III (note the inverted x-axis). The capacitance shows a first peak when $\mu$ crosses the upper valence bands, which consist of a heavy hole ($m_\text{hh} = 0.51m_0$) and a light hole band ($m_\text{lh} = 0.082m_0$). As $\mu$ drops below the split-off band ($m_\text{so} = 0.15m_0$), a second peak in $C_\text{tot}$ appears. The device parameters are the same as used in Fig.~\ref{fig:charges}.}
    \label{fig:bandstructures}
\end{figure}

So far we considered a single valence and conduction band, but in general multiple bands can be present. We can numerically evaluate $C_\text{tot}$ in such cases by using the DOS

\begin{align}
    g_\text{tot}(E) = \sum_{n=1}^N \Theta (E-E_n)\, g_n(E),
\end{align}

where $\Theta (E-E_n)$ is the Heaviside step function, and $g_n(E)$ is defined as

\begin{align}
    g_n(E) = \frac{g_{v,n}\sqrt{2m_n}}{\hbar\pi}\frac{1+2\alpha (E-E_n)}{\sqrt{(E-E_n)(1+\alpha (E-E_n))}}.
\end{align}

Here $m_n$ and $E_n$ the effective electron (hole) mass and the band minimum of the $n$-th conduction (valence) band, and $N$ is the total number of bands. As an example we evaluate $C_\text{tot}$ in the case a GaAs nanowire is used to host the 1DEG. We consider the heavy hole (hh), light hole (lh) and split-off (so) valence bands~\cite{vurgaftman2001band}, as indicated in the inset in Fig.~\ref{fig:bandstructures}. This figure shows the resulting effective capacitance, and exhibits two peaks due to the structure of the valence bands. The first peak corresponds to $\mu$ crossing the heavy hole and light hole bands, and the second peak to $\mu$ crossing the split-off band. Due to the lower DOS in the split-off band compared to the heavy and light hole bands, the nanowire fills with holes more quickly. This causes the capacitance to drop back to the linear capacitance sooner after the second peak, thus making that peak thinner than the first peak.

\section{Outlook} 
\label{applications}

\begin{figure}[t]
    \centering
    \includegraphics[width=\linewidth]{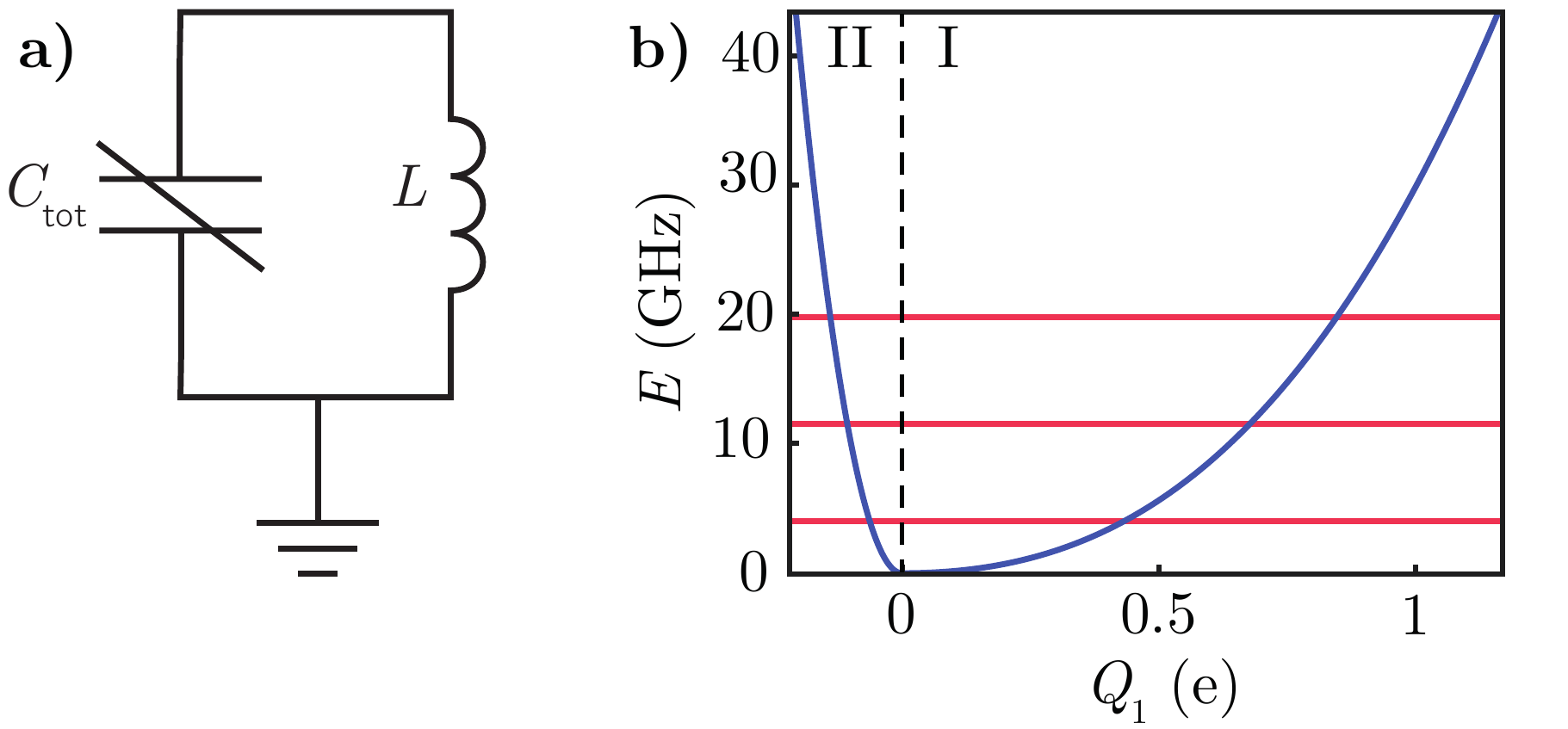}
    \caption{\textbf{a)} Schematic of the nonlinear LC-oscillator. The nanowire and lower plate of the capacitor are connected to a ground plane which allows electrons to flow to and from them. \textbf{b)} Potential energy as a function of $Q_1$. The horizontal red lines indicate the energies of the lowest three energy eigenstates.}
    \label{fig:qubit}
\end{figure}

\begin{figure}[t]
    \centering
    \includegraphics[width=\linewidth]{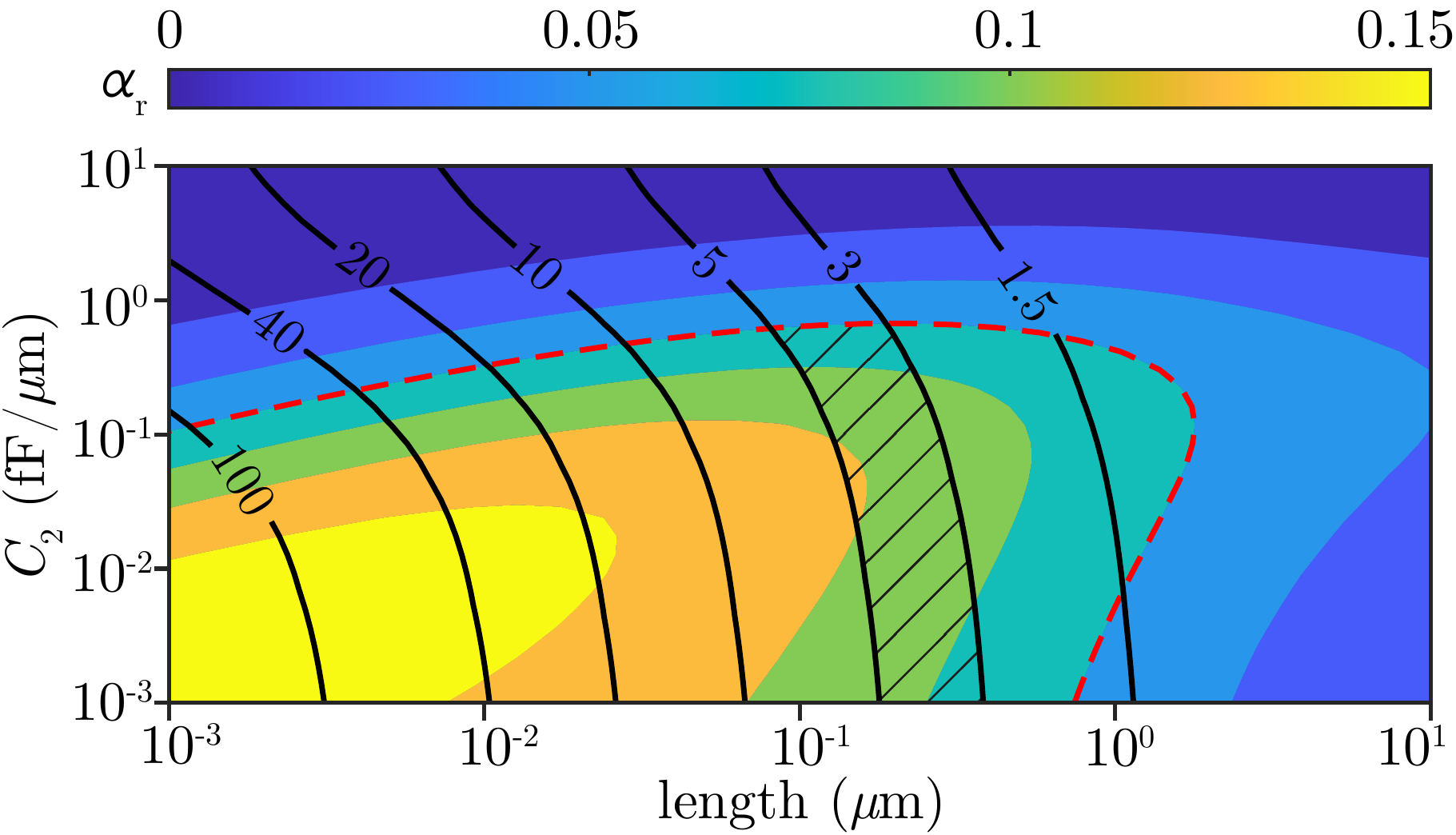}
    \caption{Operating regime for the qubit as a function of $C_2$ and the length of the nonlinear capacitor device, for $C_1 = 5.0\,$fF/$\mu$m and $L = 1000\,$nH. The black lines indicate curves of constant qubit frequency $f_{01}$, in GHz, the coloured contour plot shows the relative anharmonicity. The red dashed line indicates the line of constant anharmonicity of $\alpha_\text{r} = 0.075$. The region most useful for qubit operation is shown with the diagonal line pattern.}
    \label{fig:qubit_operation}
\end{figure}

\begin{figure}[t]
    \centering
    \includegraphics[width=\linewidth]{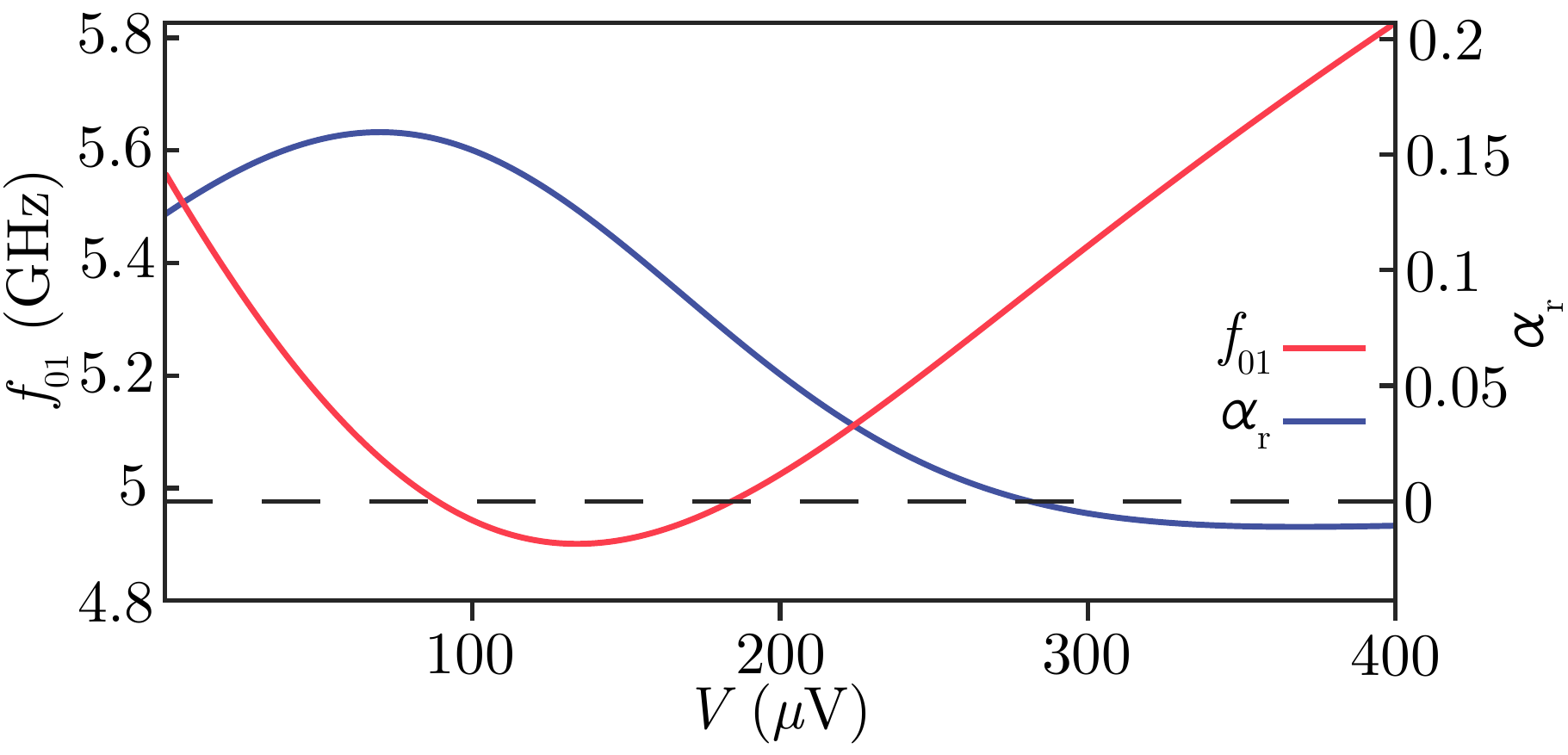}
    \caption{Qubit frequency and anharmonicity for different voltages applied to the central nanowire. The device parameters are the same as described in the text, $C_1 = 5\,$fF/$\mu$m, $C_2 = 0.15\,$fF/$\mu$m, $L = 1000\,$nH and a device length of $100\,$nm.}
    \label{fig:Voltage}
\end{figure}

By connecting the plate P1 to the reference plane by means of an inductor, as shown in Fig.~\ref{fig:qubit}a, Cooper pairs are allowed to move between the two capacitor plates. This circuit forms a nonlinear LC-oscillator which can be used to implement a qubit. The circuit is described by the Hamiltonian

\begin{equation}
    H = \frac{\phi^2}{2L} + E_\text{tot}(Q_1),
\end{equation}

with $L$ the value of the inductance and $\phi$ the conjugate momentum to the charge $Q_1$, which can also be interpreted as the magnetic flux threading the inductor loops. This Hamiltonian is derived in Appendix~\ref{qubit}. The length of the nonlinear capacitor, the geometric capacitances $C_1, C_2$ and the inductance $L$ are four design parameters which can be used to optimize the qubit operation. Fig.~\ref{fig:qubit_operation} shows the qubit transition frequency $f_{01}$ and the relative anharmonicity $\alpha_\text{r}$ between the first three energy levels, $\alpha_\text{r} = (f_{12} - f_{01})/f_{01}$, where $h f_{mn}$ is the energy difference between the levels $m$ and $n$, and $h$ is Planck's constant. The black lines are lines of equal frequency, and the coloured contour plot shows the relative anharmonicity. Currently, the most commonly used qubits in superconducting quantum processors are transmons, which have qubit frequencies of around $5$GHz and a relative anharmonicity of -0.06 and stronger~\cite{zhang2022high}\cite{kjaergaard2020superconducting}. This operation region for the NLC-qubit is indicated in Fig.~\ref{fig:qubit_operation} as the black-striped region, with $3$GHz $<f_{01}<5$GHz, and $\alpha_\text{r} > 0.075$. Outside of this region the NLC-qubit can obtain even greater anharmonicities. Apart from the operation parameters, there are a few key differences with the transmon. Firstly, the anharmonicity of the NLC-qubit described here is positive, as opposed to the negative anharmonicity of the transmon. Secondly, $f_{01}$ and $\alpha_\text{r}$ are electronically tunable with a voltage $V_\text{q}$ on the nanowire, instead of the magnetic tunability of the transmon, as shown in Fig.~\ref{fig:Voltage}. The applied voltage shifts the Fermi level in the nanowire upwards, causing the boundary between regimes I and II to shift to the value $Q_1 = -C_2V_\text{q}$. Due to this shift, the anharmonicity shows a clear maximum, as at low positive values of $V_\text{q}$ the nonlinear behaviour of the capacitor extends to negative values of $Q_1$, increasing $\alpha_\text{r}$. Increasing $V_\text{q}$ to larger positive values results in a decrease of the nonlinearity which eventually converges to $\alpha_\text{r} = 0$, since the capacitance converges to the constant value of $C_1C_2/(C_1+C_2)$.
This tunability of the anharmonicity suggests the nonlinear capacitor can also be used in an electrically tunable coupler. The higer-order terms in Eq.~(\ref{eq:totE}) [see also Appendix~\ref{qubit}, Eq.~(\ref{energy})] can induce couplings between higher qubit levels, making it possible to excite the third level in a qubit directly.

\section{Conclusions}

We presented an analytical and numerical study of a nonlinear capacitor composed of two metallic electrodes with a nanowire acting as a 1DEG in between. Due to the low density of states in the nanowire, a quantum capacitance is formed which depends on the voltage over the plates, resulting in a nonlinear total capacitance. We identified three different capacitive regimes, one linear and two nonlinear, depending on the position of the Fermi level in the nanowire. A numerical investigation into the effects of non-zero temperatures, non-parabolic energy bands and additional conduction bands was performed, and their influence on the capacitance was explained. We found temperatures up to $1$K to have a negligible effect on the capacitance, although thermal noise and other noise sources were not considered.
Owing to the fact that no large magnetic or electric fields are needed to obtain a nonlinear behaviour, this setup can be used for applications in cQED, such as the implementation of a qubit. This qubit could achieve similar transition frequencies and anharmonicities as the often used transmon, but with additional features such as a positive anharmonicity and electronic control of the energy levels.

\section{Acknowledgements}
This work was supported by imec’s Industrial
Affiliation Program.

\appendix
\section{Geometric capacitance} 
\label{geom_cap}

\begin{figure}
    \centering
    \includegraphics[width=\linewidth]{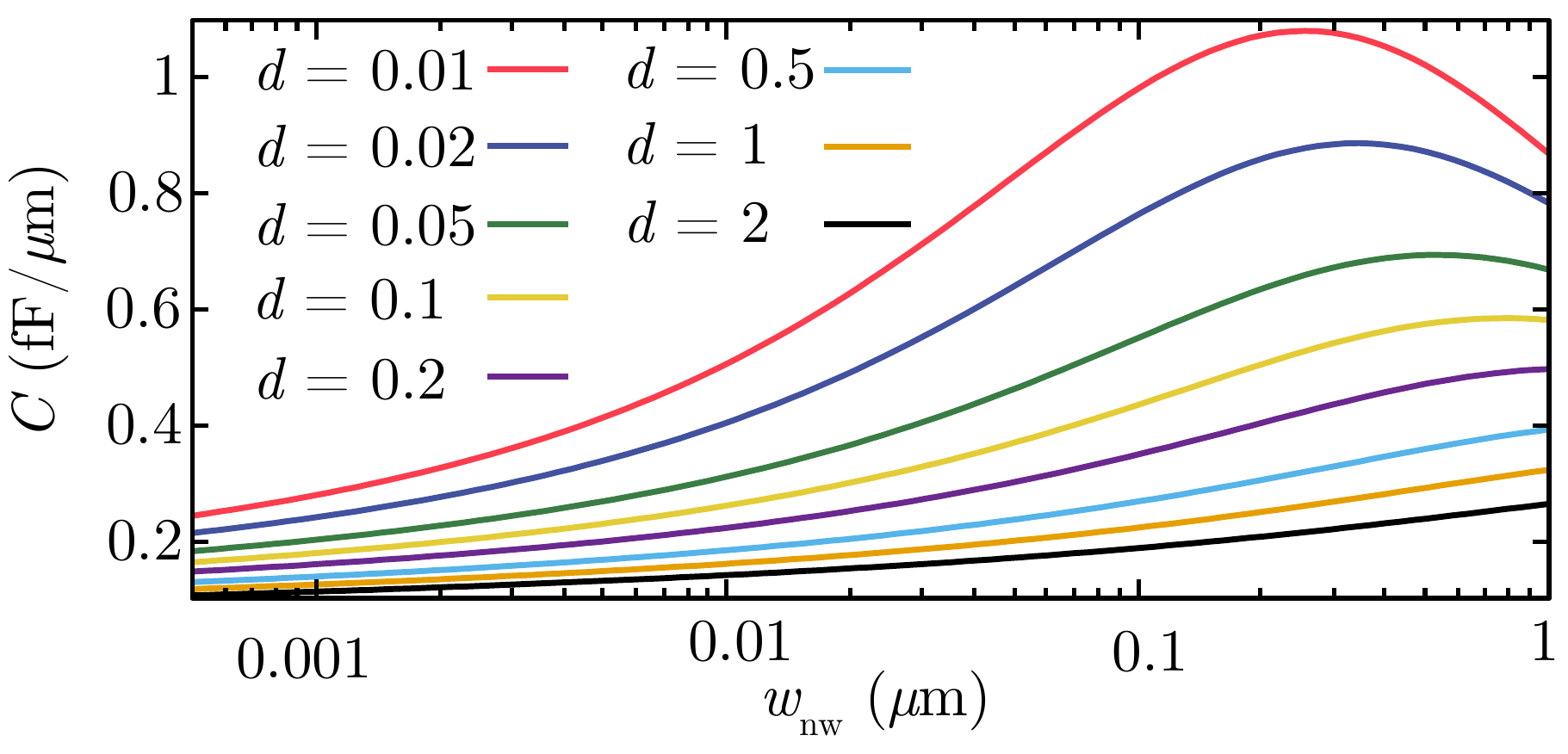}
    \caption{Capacitance as a function of the width of the nanowire for different distances between the plate and the nanowire (in units of $\mu$m). For this we have considered a coplanar geometry with a metallic plate with a width $w_\text{pl}=1\,\mu$m. We have furthermore considered Si as the dielectric between the plates, with $\varepsilon = 11.68$.}
    \label{fig:cap_geom}
\end{figure}

In the case of a coplanar capacitor, the geometric capacitances per unit of length $C_1$ and $C_2$ depend on the geometry of the system~\cite{Zypman2019}, i.e., the length of the capacitor, the width of the plates and the relative permittivity of the dielectric. In order to estimate the range of the capacitances for realistic geometries, we evaluate the capacitance between a superconducting plate and a quasi-one-dimensional electron gas in the same coplanar geometry as in Fig.~\ref{fig:device}a. The results, shown in Fig.~\ref{fig:cap_geom} are obtained by numerically solving the integral equation

\begin{align} \label{eq:int_eq}
    V(\mathbf{r'}) ={}&{} \frac{1}{4 \pi \varepsilon} \int_A \frac{\sigma (\mathbf{r'}) d\mathbf{r'}}{|\mathbf{r}-\mathbf{r'}|},
\end{align}

considering the quasi-one-dimensional electron gas as a narrow metallic plate with width $w_\text{nw}$. In Eq.~\ref{eq:int_eq}, $\sigma (\mathbf{r})$ is the charge distribution in the capacitor plates due to a potential $V(\mathbf{r})$, $\varepsilon$ is the dielectric constant, and the integration domain is the two-dimensional plane of the coplanar capacitor, with the constraint that the charge density is only non-zero within the two metallic plates. After imposing a constant electric potential $V$ in the plates, we solve the equation above and find the charge distribution $\sigma (\mathbf{r})$ as a function of the width of each plate and the distance between them~\cite{Paul2008,Pollack2002}.

\section{Qubit derivation} 
\label{qubit}

\begin{figure}
\centering   \includegraphics[width=\linewidth]{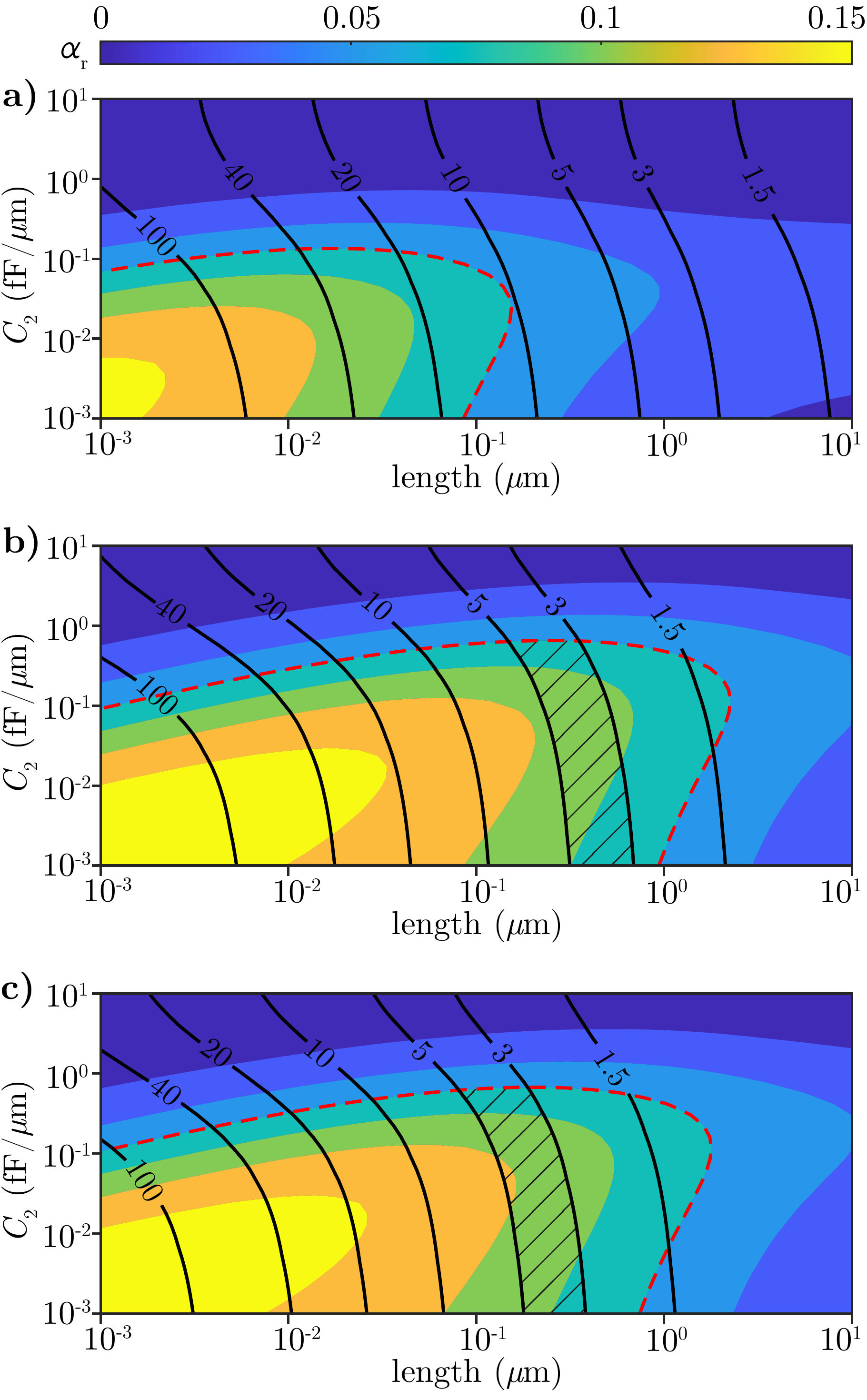}
   \label{fig:Ng1} 
\caption[Plots]{Relative anharmonicity and frequency of the qubit for a range of device lengths and capacitances ($C_2$). The coloured contour plot shows the relative anharmonicity, the black contour lines are lines of constant $f_\text{01}$ in GHz. The red line indicates the points where $\alpha_\text{r} = 0.075$. \textbf{a)} For a qubit with a frequency of $5GHz$, with a capacitance $C_1 =1\,$fF/$\mu$m and an inductance $L = 500\,$nH, we observe that the maximal anharmonicity is around $0.05$. \textbf{b)} With $C_1 = 5\,$fF/$\mu$m, the region with $\alpha_\text{r} \leq 0.075$ increases, while the frequency contour lines are shifted to the left (towards smaller device lengths). \textbf{c)} The same as \textbf{b)} but with an inductance $L = 1000\,$nH. This increase in the inductance leaves region of large anharmonicity mostly unchanged, but pushes the frequencies further to the left. In the striped region, the frequency and anharmonicity are similar to those of a typical transmon qubit~\cite{krantz2019quantum}.}
\label{fig:anharmonicity}
\end{figure}

Here we consider a nonlinear capacitor connected to an inductor, as shown in Fig.~\ref{fig:qubit}. As main coordinate of the system we use the charge $Q_1$ accumulated on the upper capacitor plate P1. As the upper capacitor plate is only connected to the inductor, the current through the inductor is the rate at which P1 charges and discharges, $I_L = \dot{Q}_1$. The lagrangian of the system can thus be written as

\begin{equation}
    \mathcal{L} = \frac{L}{2}\dot{Q}^2_1 - E_\text{C}(Q_1),
\end{equation}

With $E_\text{C}(Q_1)$ the analytical expression of the capacitor energy as derived in section~\ref{analytical model}. This expression reads

\begin{equation}
    E_\text{C}(Q_1) = \begin{cases} E_\text{II}(Q_1) & \text{if } Q_1 \leq C_2V_q/e\\
   E_\text{I}(Q_1) & \text{if } Q_1 > C_2V_q/e
    \end{cases}
\end{equation}

with

\begin{equation}
    E_\text{II}(Q_1)=\left( \frac{1}{2 C_1} + \frac{1}{2 C_2} \right) Q_1^2,
\end{equation}

the capacitance in the linear regime, and 

\begin{align}
    E_\text{I}(Q_1) ={}&{} \left( \frac{1}{2 C_1} + \frac{1}{2 C_2} \right) Q_1^2  \notag \\
     {}&{} + \frac{4 g_0^4 e^6}{3 C_2^3} \left[ 1 + \frac{3}{2} \frac{C_2 (Q_1 + C_2 V_q)}{g_0^2 e^3} \right] \notag \\
     {}&{} - \frac{4 g_0^4 e^6}{3 C_2^3} \left[ 1 + \frac{C_2 (Q_1 + C_2 V_q)}{g_0^2 e^3} \right]^{3/2},
     \label{energy}
\end{align}

the capacitance in the nonlinear regime. The momentum conjugate to $Q_1$ is then $\frac{\partial \mathcal{L}}{\partial \dot{Q}_1} = L\dot{Q}_1 = LI_L = \phi$, with $\phi$ the magnetic flux through the inductor loop. After performing a Legendre transform, we obtain the Hamiltonian

\begin{equation}
    H = \dot{Q}_1\phi - \mathcal{L} = E_C(Q_1) + \frac{1}{2L}\phi^2,
    \label{eq:Hamiltonian}
\end{equation}

which we diagonalize numerically. The resulting energy spectrum is shown in Fig.~\ref{fig:qubit}, where the potential energy clearly deviates from parabolicity, resulting in an anharmonic energy spectrum. The relative anharmonicity $\alpha_\text{r}$ is defined as $\alpha_\text{r} = \frac{f_{12}-f_{01}}{f_{01}}$ with $f_{mn}$ the transition frequency between levels $m$ and $n$.  As shown in Fig.~\ref{fig:anharmonicity}, both $f_\text{01}$ and $\alpha_r$ depend strongly on the device parameters $C_1$, $C_2$, $L$, and the length of the nonlinear capacitor. To obtain a high anharmonicity during fabrication, $C_1$ should be made as large as possible, as increasing $C_1$ generally raises the anharmonicity surface. This can be seen in the difference between  Fig.~\ref{fig:anharmonicity}a and Fig.~\ref{fig:anharmonicity}b, where $C_1$ is increased from $1\, $fF/$\mu$m to $5\, $fF/$\mu$m. The inductance can then be used to set $f_{01}$. Comparing Fig.~\ref{fig:anharmonicity}(b) with Fig.~\ref{fig:anharmonicity}c, we see that increasing the inductance from $500\,$nH to $1000\,$nH shifts the lines of equal frequency towards the lower device lengths. Fig.~\ref{fig:anharmonicity}c shows a region (marked with a striped pattern) where the qubit frequency is below $5\,$GHz, and the anharmonicity is higher than $\alpha_\text{r} = 0.075$, which is comparable to the parameters of transmon qubits~\cite{krantz2019quantum}.

\end{document}